\documentclass[twocolumn,showpacs,preprintnumbers,amsmath,amssymb,superscriptaddress]{revtex4}
\usepackage{graphicx}% Include figure files
\usepackage{dcolumn}% Align table columns on decimal point
\usepackage{bm}% bold math

\begin{document}

\bibliographystyle{prsty}

\preprint{APS/123-QED}

\title{Fabrication of magnetic atom chips based on FePt}
\author{Y.~T. Xing}
\altaffiliation[Present address: ]{Centro Brasileiro de Pesquisas
Fisicas (CBPF), Rua Dr. Xavier Sigaud 150, Urca Rio de Janeiro CEP
22290-180, Brazil}

\author{I. Barb}
\author{R. Gerritsma}
\author{R.~J.~C. Spreeuw}
\author{H. Luigjes}
\affiliation{Van der Waals-Zeeman Institute, University of Amsterdam, \\
Valckenierstraat 65, 1018 XE Amsterdam, The Netherlands}
\homepage{http://www.science.uva.nl/research/aplp/}

\author{Q.~F. Xiao}
\altaffiliation[Present address: ]{Department of Physics and
Astronomy, University of Victoria, Victoria, British Columbia,
Canada}

\author{C. R\'{e}tif}
\affiliation{FOM-Institute of Atomic and
Molecular Physics, Kruislaan 407, PO Box 41883, NL-1009 DB,
Amsterdam, The Netherlands}

\author{J.~B. Goedkoop}
\email{goedkoop@science.uva.nl}
\affiliation{Van der Waals-Zeeman Institute, University of Amsterdam, \\
Valckenierstraat 65, 1018 XE Amsterdam, The Netherlands}
\homepage{http://www.science.uva.nl/research/aplp/}

\date{\today}% It is always \today, today,
             %  but any date may be explicitly specified

\begin{abstract}
We describe the design and fabrication of novel all-magnetic atom
chips for use in ultracold atom trapping. The considerations
leading to the choice of nanocrystalline exchange coupled FePt as
best material are discussed. Using stray field calculations, we
designed patterns that function as magnetic atom traps. These
patterns were realized by spark erosion of FePt foil and e-beam
lithography of FePt film. A mirror magneto-optical trap (MMOT) was
obtained using the stray field of the foil chip.
\end{abstract}

% PACS, the Physics and Astronomy
     % Classification Scheme.
%\keywords{Suggested keywords}%Use showkeys class option if keyword
                              %display desired
\maketitle

\section{\label{sec:level1}Introduction}

The rapidly proceeding development of techniques to manipulate the
motion of ultracold atoms has recently led to the invention of
so-called atom chips. An atom chip is a planar structure that
produces a designer magnetic field pattern in the vacuum above the
structure. The magnetic field minima are used to confine atoms
carrying a magnetic dipole moment {$\bf{\mu}$}. The confining
force results from the magnetic dipole interaction (-{$\bf{\mu
\cdot B}$). An endless variety of structures can be created,
including atomic wave guides, interferometers, arrays of atom
traps for quantum information processing and atomic conveyor belts
\cite{FolKruHen02,Rei02}. Atom chips are now emerging as an
extremely powerful and versatile tool
\cite{FolKruHen02,ReiHanHan99}, and even seem to provide a simpler
method to achieve Bose-Einstein condensation (BEC)
\cite{HanHomHan01,OttForSch01,SinCurHin0503619}. Furthermore, atom
chips allow one to reduce the time scale of trapping experiments,
which could imply that the vacuum requirements can be relaxed
\cite{ReichelSA05}.

So far, atom chips use planar patterns of current-carrying wires
to generate the magnetic fields. Here we investigate a promising
alternative based on hard magnetic materials patterned by
micromachining or lithographic techniques
\cite{MLod04,Dav99,SidMcLHan02,SinRetHin0502073}.

The use of hard magnetic films has many potential advantages.
Firstly, since no lead wires are required, magnetic structures
offer a great design flexibility, allowing for more intricate
patterns including structures that are topologically impossible
with current-carrying wires. Secondly, there is no resistive power
dissipation, current noise from power supplies nor stray field
from lead wires. Thirdly, the conductivity of magnetic materials
can be orders of magnitude lower than Cu or Au. This could help to
overcome the lifetime reduction due to coupling of the cold atom
cloud to random thermal currents in the chip structure
\cite{HenPotWil99,JonValHin03,HarMcGCor03}. Finally, we calculate
that magnetic field gradients of $\thicksim$ 20 kT/m should be
achievable, considerably higher than the highest values that have
been achieved with current conductors.
\\
An obvious advantage of current-conducting chips is the
possibility to switch or modulate the currents. In magnetic chips
the magnetic field pattern can only be manipulated by applying
fields generated by external electromagnets. However, hybrid
chips, combining the best parts of the two techniques, are
technologically possible and will be used in future.
\\
Currently, we are exploring the possibilities of planar magnetic
structures for the confinement and manipulation of cold atoms in
all-magnetic atom chips \cite{BarGerSpr05}. In this paper, we
discuss aspects of the fabrication of such chips. In particular,
we discuss the material requirements that led us to the choice of
FePt as most suitable material. We give details on the fabrication
of two FePt-based atom trap designs of respectively sub-millimeter
and sub-micron dimensions. The first design consists of a pattern
cut out of a 40 $\mu$m thick FePt foil using spark erosion. The
stray field of this structure allowed the operation of a
millimeter-size magneto-optical trap for Rb atoms, thus showing
the suitability of FePt as a material. The second design is an
array of micron sized magnetic traps patterned in a 250 nm thick
FePt film on Si using e-beam lithography techniques. We describe
details on the FePt film optimization and the patterning process
for this design.

\section{Selection of the hard magnetic material}

In permanent magnetic atom chips the cold atoms are statically
trapped in the stray field generated by permanent magnetic
structures which, when properly designed, do not require external
fields. The stray field determines the depth of the available atom
traps which should be on the order of 0.1-1 mT.
\\
The first material requirement is obviously a high magnetization
in order to effectively generate the strong field gradients.
However, each magnetic material tends to break up in domains
exactly in order to reduce the stray field that it produces. This
demagnetization problem is compounded by the external fields used
for loading the atoms into the traps. A second requirement is
therefore that the material has a high coercivity.

    \begin{figure}[tbp]
    \centering{\includegraphics[width=60mm]{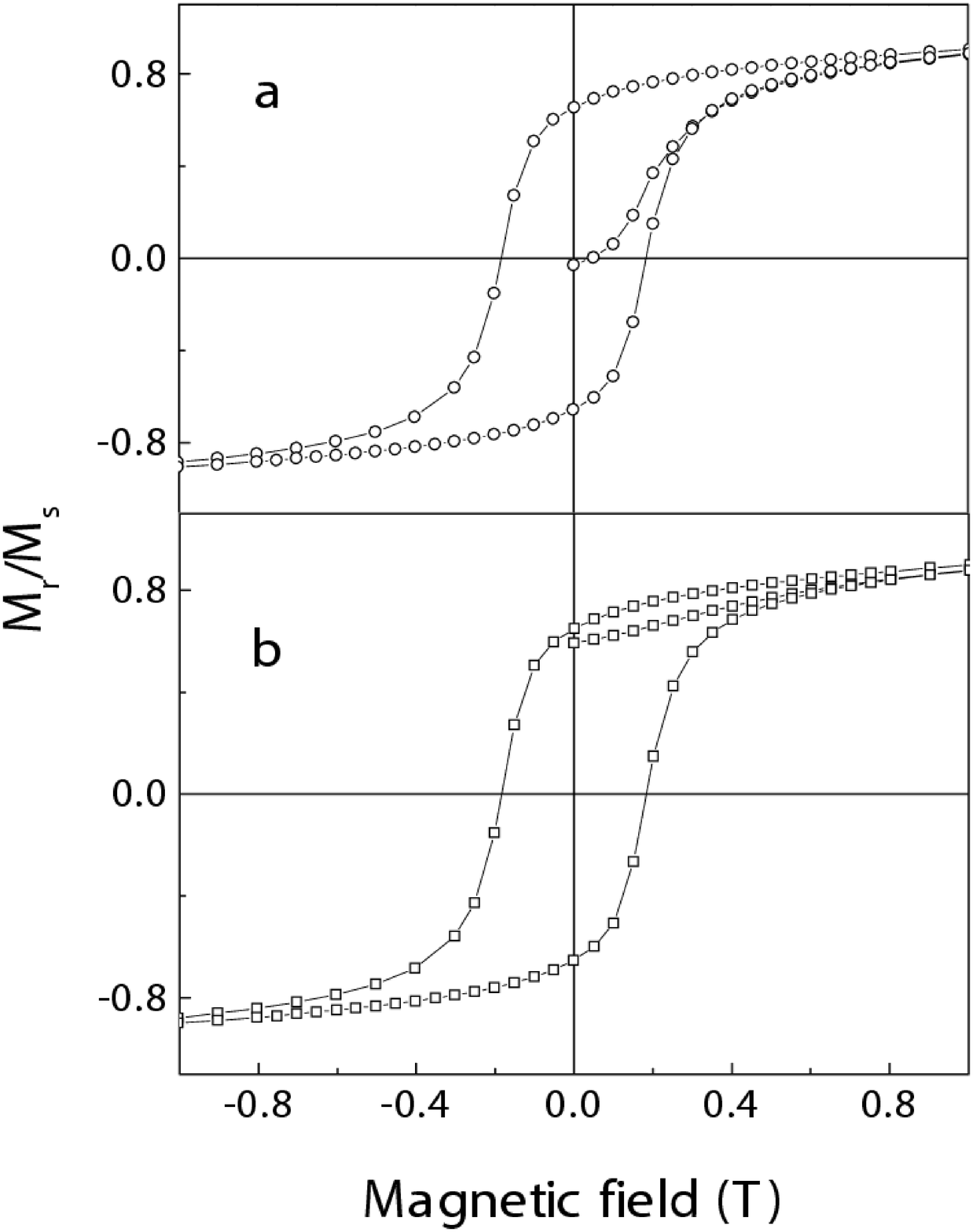}}
    \caption{Magnetization loops of a 40 $\mu$m foil of nanocrystalline \\
    FePt with in-plane magnetization, before and after baking, (a) and
    (b), at 170 $^{\circ}$C for 24 hours.} \label{fig1}
    \end{figure}

Since the most exciting prospect for atom chips is the possibility
to integrate many microscopic atom optical devices on a single
chip, the material should be suitable for preparation as a film.
Clearly, the magnetic fields produced by such a thin film
structure scales with its thickness. In order to produce a large
enough stray field at a distance of a few $\mu$m from the surface
with a few micron line width, the material thickness should be in
the range of 100 nm to 1 $\mu$m.
\\
In addition, the material should be corrosion resistant in order
to allow micromachining or lithographic patterning. In order to
avoid uncontrolled spatial variations of the stray field, the
material must also be highly homogeneous, which puts severe
constraints on these manufacturing processes. Finally, the traps
are operated in ultrahigh vacuum, which means that the
magnetization should survive a 24 hours bake out at
150$^{\circ}$C.
\\
The combination of high magnetization and high anisotropy limits
the range of materials naturally to the strongest room temperature
permanent magnets which are Nd$_{2}$Fe$_{14}$B, Co$_{5}$Sm and
FePt alloys (see e.g. \cite{WelDoe00}). Among these, the hard
magnets Co$_{5}$Sm and Nd$_{2}$Fe$_{14}$B have excellent magnetic
properties, but are difficult to grow as thin films. Moreover,
Nd$_{2}$Fe$_{14}$B is unstable at bake out temperatures. The other
candidates are the CoPt and FePt systems, where the latter has the
higher magnetization and was therefore selected as the best
material. FePt has been studied extensively both in bulk
\cite{MXiaBruBus04,MThaBruBoe02} and thin film
\cite{MKimShiKan02,MPlaWieLau02,MWeiSchFah04,MBerAttSam03,MYanPowSel03}
form since it combines high magneto-crystalline anisotropy with
high saturation magnetization M$_{s}$ \cite{WelDoe00} and has
excellent stability and corrosion resistance. FePt has a
disordered face-centered cubic (fcc) structure at high
temperature, which has a very high M$_{s}$ but is magnetically
soft. The low temperature equilibrium structure on the other hand
is face-centered tetragonal (fct or L1$_{0}$), in which the Fe and
Pt order in an atomic multilayered structure with stacking in the
[111] direction. This phase has a lower M$_{s}$ but very high
magneto-crystalline anisotropy and coercivity. It has been shown
that annealing of either the fcc phase obtained from the melt
\cite{MLiuLuoSel98} or as-deposited thin films produces
nanocrystalline composites of the two phases in which the
nanocrystallites of the hard fct phase orient the surrounding soft
phase by exchange coupling. This results in a material which
combines high magnetization with isotropic hard magnetic behavior.

\section{Fabrication of an atom chip based on $FePt$ foil}

As a first step towards thin film chips we created a trap from
FePt foil produced by bulk metallurgic techniques
\cite{MLiuLuoSel98}. This method is suitable for traps with
millimeter dimensions. A Fe$_{0.6}$Pt$_{0.4}$ alloy was made by
arc-melting in a purified Ar atmosphere. The melt was cast into a
water-cooled copper mould to get cylindrical samples with a
diameter of 1.5 mm. These were sealed into evacuated quartz tubes
and homogenized at 1300 $^{\circ}$C for 3 hours after which they
were quenched into ice water without breaking the quartz tube to
produce fct grains in an fcc matrix. Subsequently the samples were
annealed for 16 minutes at 580 $^{\circ}$C in order obtain the
nano-composite FePt.

    \begin{figure}[t]
    \centerline{\includegraphics[width=70mm]{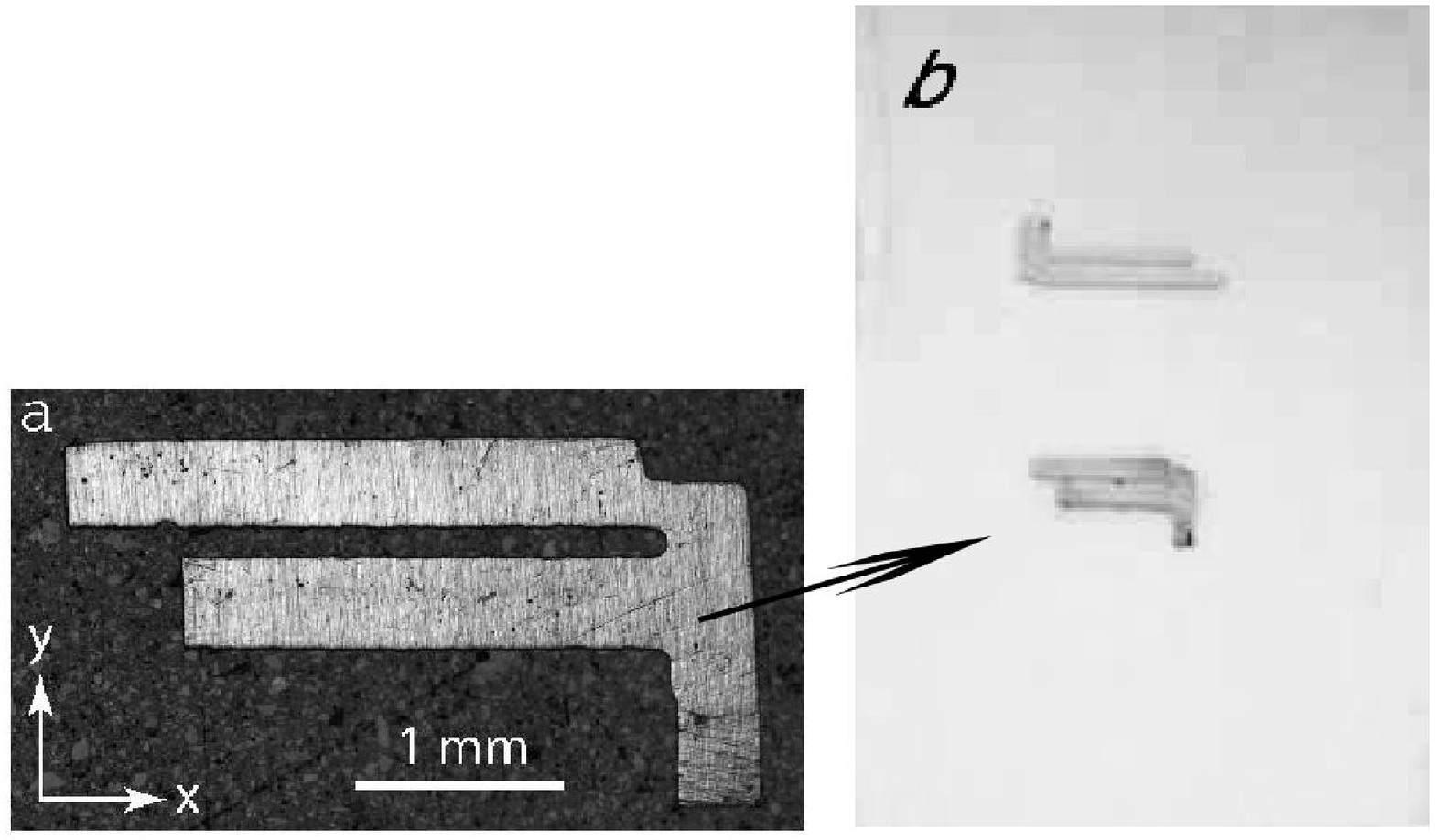}}
    \vspace*{5mm}
    \centerline{\includegraphics[width=40mm]{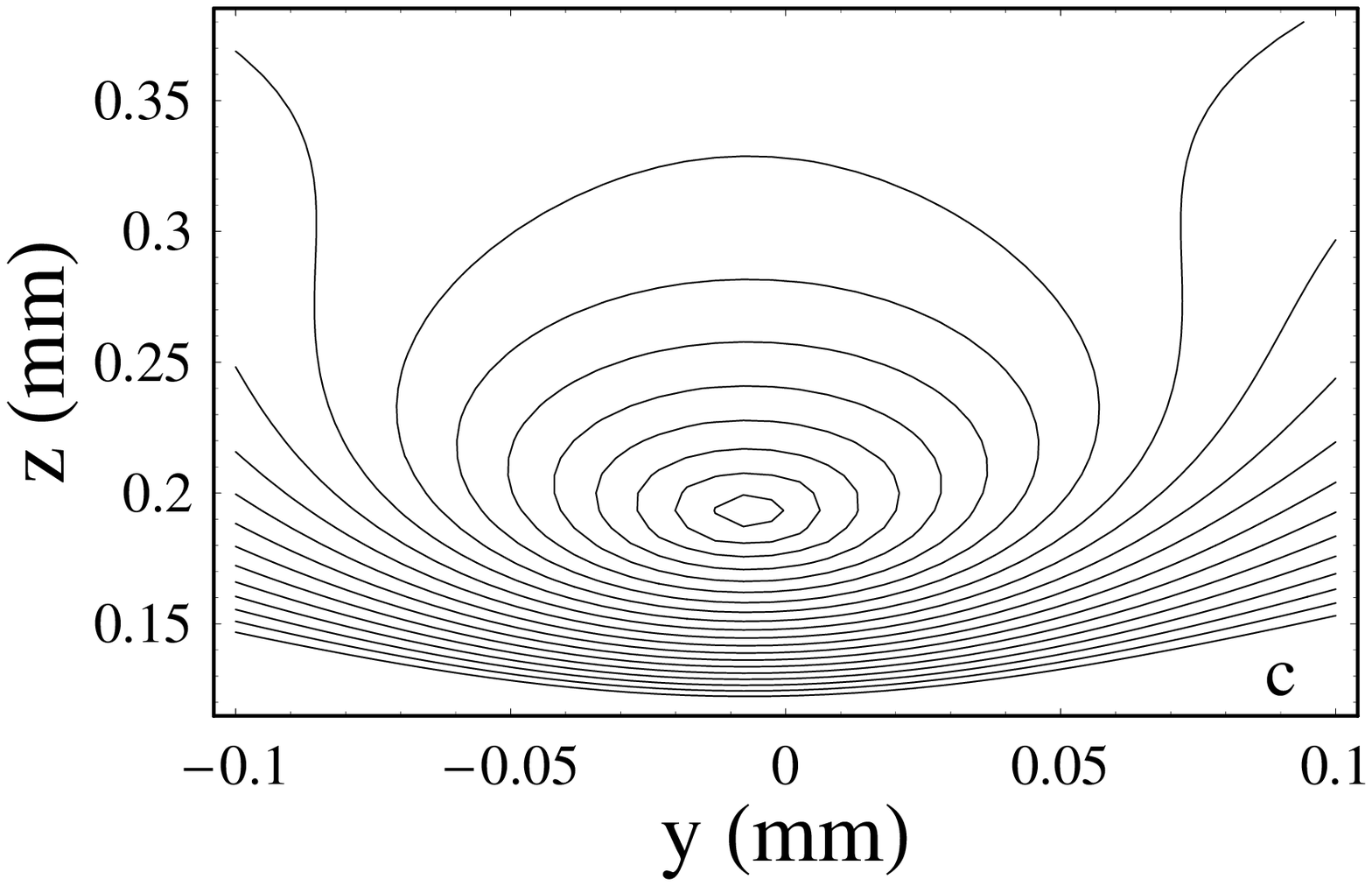}\includegraphics[width=40mm]{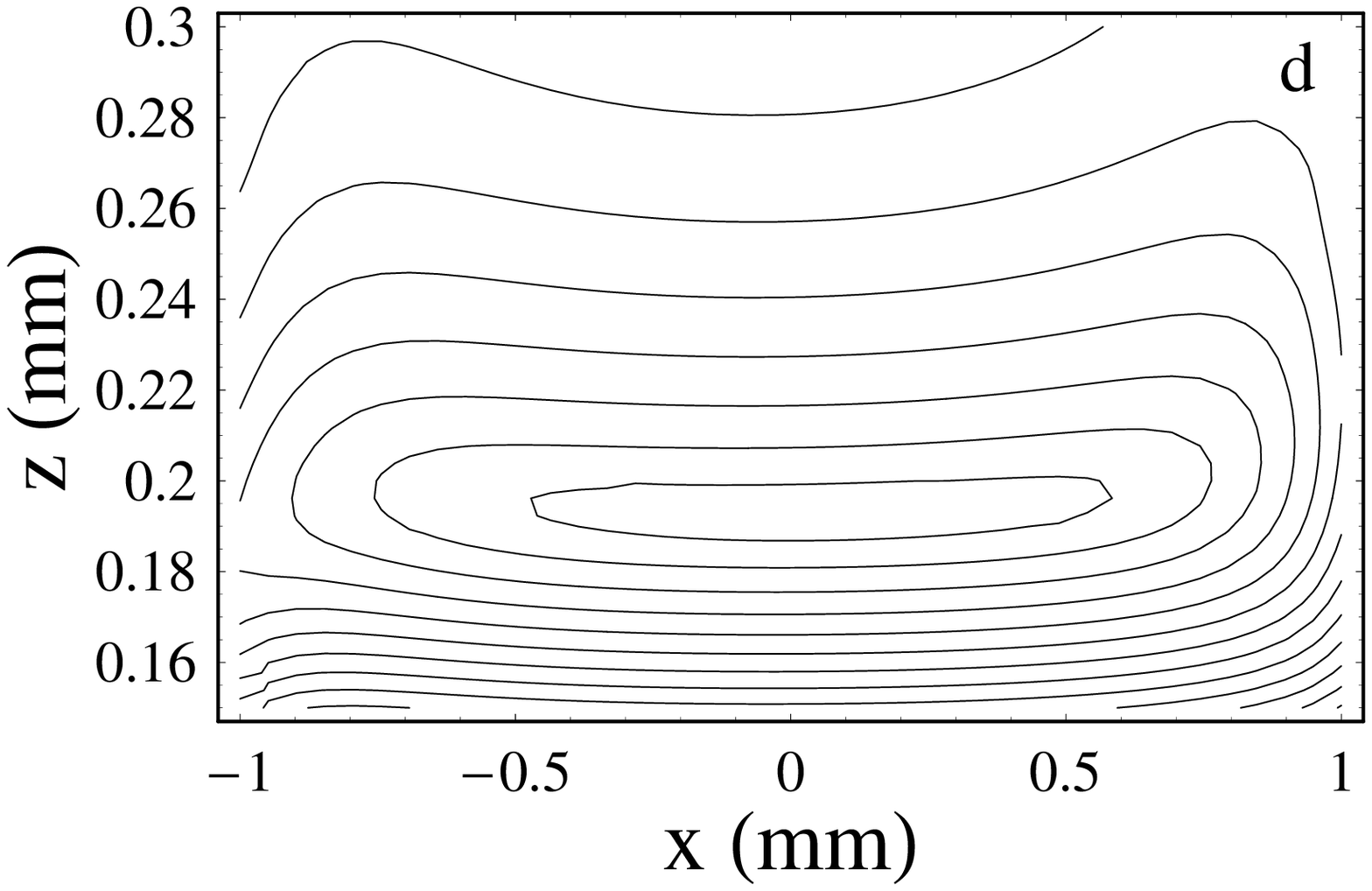}}
    \caption{(a) Foil atom chip. (b) Photograph of two
    FePt structures with slightly different dimensions glued
    on an aluminum mirror. (c) $y$-$z$ and (d) $x$-$z$ plots of
    the calculated magnetic stray field for the lower structure
    in (b). Shown are contours of equal magnitude of magnetic field B
    (i.e. equipotential lines), with a spacing of 6 G between the
    contours, in planes containing the trap minimum. The chip
    surface is at z=0.2 mm.} \label{fig2}
    \end{figure}

A bar of this material was rolled into a 100 $\mu$m thick foil and
then mechanically polished to 40 $\mu$m thickness. Figure 1a shows
the in-plane magnetization loop of this material measured with
SQUID. From this figure one can see that the ratio of remanent to
saturation magnetization M$_{r}$/M$_{s}$ is about 0.8, the
reduction being due to canting of magnetization in grains which
have their magnetic anisotropy directions away from the field
direction. The coercivity is about 0.2 T, large compared to the
external field that is applied to manipulate the atoms (10 mT).
Since saturation requires at least 3 Tesla it is impossible to
magnetize in-situ. It is therefore crucial that the material
maintains its magnetization during the 150 $^{\circ}$C vacuum
bake-out. As figure 1b shows, the magnetization of a saturated
sample that was baked at 170 $^{\circ}$C for 24 hours decreased
less than 5 $\%$.

CNC controlled spark erosion using a 50 $\mu$m wire was used to
produce elongated F-like patterns (Fig.2a). After cutting, the
damaged surface layer of the outer edges of the F-shape sample was
removed by mechanical polishing. Two such samples were mounted on
an aluminum mirror using UHV compatible glue, as shown in Fig. 2.
Finally, the samples were magnetized in a 3 Tesla field oriented
along the stem of the F ($y$-direction) before they were put into
the vacuum chamber. The two arms of this F structure form in-plane
dipole magnets in the $y$,$z$-plane with the field in the short
direction, which combined produce a zero field line above the gap
between them. The stray field from the stem of the F has a
component in the $x$-direction which serves to offset this minimum
in order to avoid Majorana spin flips
\cite{GotIofTel62,Pri83,BerEreMet87,BagLafPri87}.

The structure was designed as one piece to obtain sufficient
mechanical precision. The smallest size of the gap between the two
hands of the big F shape is determined by the diameter of the wire
that we used to cut the sample. In the optimization we only used
one gap size and tuned the shape and the dimensions of the pattern
according to the gap size and the thickness of the foil.

The calculated stray field in the $y$, $z$ and $x$, $z$ planes of
the F pattern of Fig. 2a are shown in Fig. 2c and 2d. The figure
shows contour lines of equal magnitude of the magnetic field B =
$|\mathbf{B}|$, which effectively acts as the potential energy for
cold atoms. The calculation yields harmonic trapping frequencies
in the bottom of the trap of 51~Hz and 6.8~kHz, in the
longitudinal and transverse directions, respectively. The trap
depth is about 11 G. The upper F pattern in Fig. 2b has calculated
trapping frequencies of 34~Hz and 11~kHz and a trap depth of 33~G.
More details about the trap design will be published in a separate
publication \cite{BarGerSpr05}. Figure 3 shows the fluorescence
from a cloud of $\sim$2 $\times$ 10$^{6}$ cold Rb atoms trapped
under this foil in a so-called mirror magneto-optic trap (MMOT),
using a combination of a quadrupole magnetic field and four laser
beams. It should be noted that in this case the quadrupole field
is achieved by adding a 2 G external field that cancels the field
of the F structure at a distance of 2 mm from the surface. This
creates a point of zero magnetic field, with a gradient of $\sim$
15 G/cm. The MMOT merely uses the decaying magnetic stray field
far from the structure. True magnetic trapping has also been
achieved with this structure, as described in ref.
\cite{BarGerSpr05}.

    \begin{figure}[t]
    \centerline{\includegraphics[width=60mm]{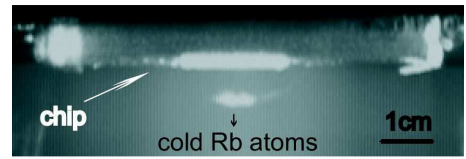}}
    \caption{Fluorescence image of a cloud of cold $^{87}$Rb atoms
    under the mirror. Mirror magneto-optical trap, using the stray
    field of the structure, at about 2 mm below the lower "F" shown in
    Fig. 2b.}\label{fig3}
    \end{figure}

\section{Atom chip based on lithographically patterned $FePt$ films}

Our second sample is based on a FePt film deposited on Si. The
thickest films that can be grown with vacuum deposition and that
can be structured using lithographic techniques are in the micron
range, implying scale reduction of a factor 40 or more with
respect to the design discussed above. Here we used 250 nm
Fe$_{50}$Pt$_{50}$ films, co-evaporated by Molecular Beam Epitaxy
(MBE) from Fe and Pt targets on rotating Si substrates at 350
$^{\circ}$C. Details on the material optimization
\cite{MXinEljGoe04} are summarized as follows. The as-deposited
samples have a strongly disordered structure, in which both fcc
and fct phase are present according to X-ray diffraction. This
film is post-annealed to obtain the proper characteristics. Figure
4 shows the magnetic properties of the thick film as a function of
annealing temperature. From this figure one can see that the best
magnetic properties (optimum M$_{r}$/M$_{s}$) are: M$_{s}$=750
kA/m, M$_{r}$/M$_{s}$ = 0.93, H$_{c}$= 0.83 T for out-of-plane
magnetization (T$_{anneal}$=450 $^{o}$C) and M$_{r}$/M$_{s}$ =
0.90, H$_{c}$= 1 T for in-plane magnetization (T$_{anneal}$=500
$^{o}$C).  These loops show that the magnetic properties of these
films are comparable to the bulk material. A very desirable aspect
of these results is that these films can be used both for in-plane
and for out-of-plane chip designs. In the remainder of this
section we will describe the production of an in-plane design on
the basis of this film material.

The films were patterned with e-beam lithography, using a hard
negative photoresist (SU-8), to save writing time. The patterning
process consists of the following steps: First the surface is
roughened using Ar plasma pre-etching in order to increase the
adhesion of the SU-8 on the FePt surface. After spin-coating and
pre-bake a 340 nm thick SU-8 photoresist layer was obtained. The
desired pattern was written in this resist using a JEOL 6460 SEM
equipped with a Raith Quantum pattern generator. After writing the
resist is post-baked, developed to remove the non-exposed resist
and hard-baked in order to harden the remaining resist layer.
\\
The resist structure is subsequently transferred to the FePt layer
by Ar plasma etching with an Oxford Plasmalab 80 Plus reactive ion
etching system. The samples were etched for 7 minutes  in order to
slightly over-etch the FePt layer to get very clear patterns.
Figure 5a shows a typical FePt pattern and the cross section of
one strip revealing a slope of the FePt edge of about 45$^{o}$,
with a roughness of about 50 nm, which is of the order of the
nanocrystalline grain size.

The etching rates of the SU-8 and FePt are almost the same (~40
nm/min) but the photoresist layer is thicker than FePt, and the
inset shows that there is still a photoresist layer left on top of
the FePt layer. It is straight forward to adjust the initial
resist thickness in order to reduce this layer. Furthermore, since
the final chip is coated with a thin Au film to obtain a highly
reflecting surface, the thin photoresist layer has little effect.

    \begin{figure}[t]
    \centerline{\includegraphics[width=60mm]{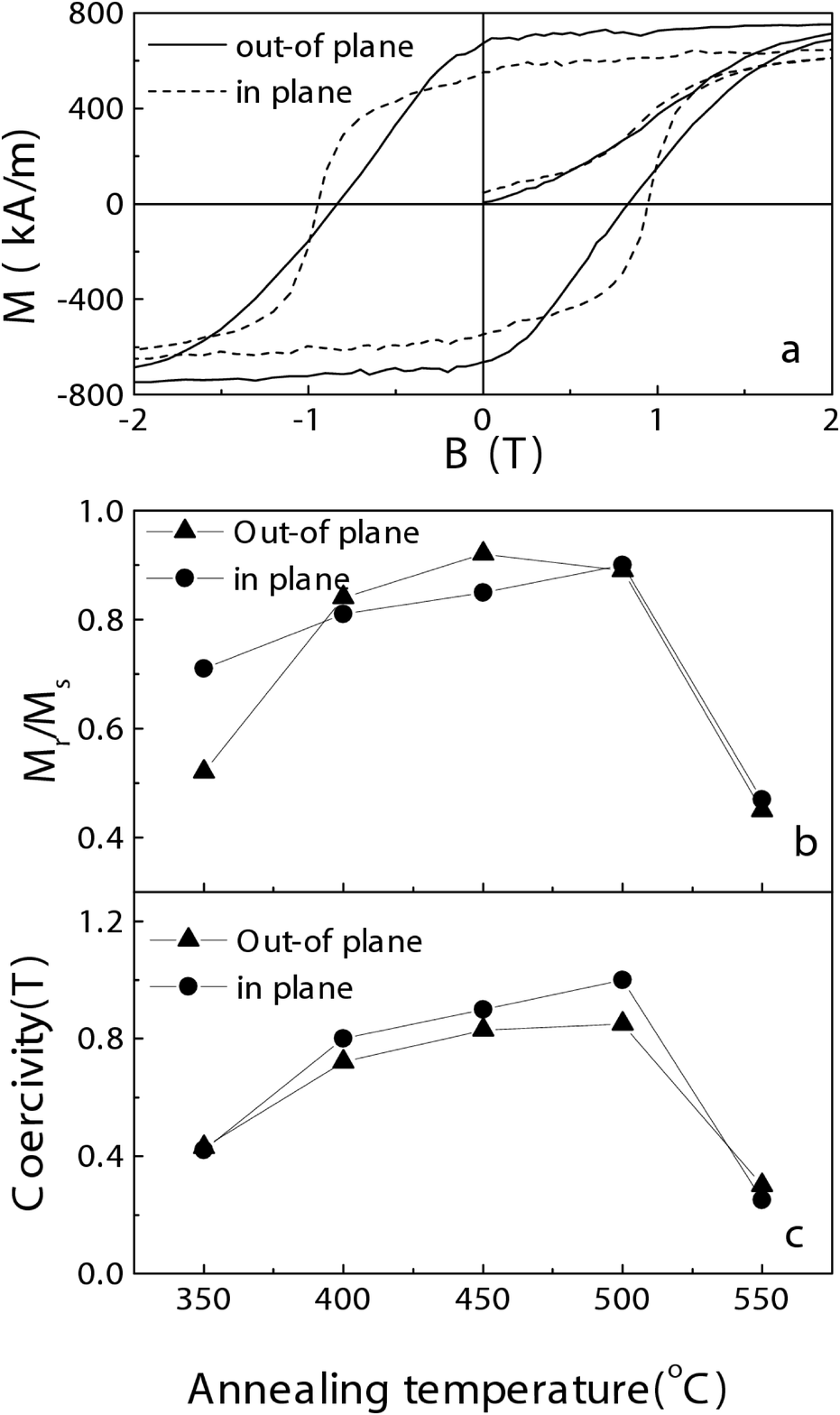}}
    \caption{Magnetic properties of the 250 nm FePt film. (a) The
    hysteresis loop after a 450$^{\circ}$C anneal for 3 minutes. (b)
    M$_{r}$/M$_{s}$ and (c) coercivity of the film as a function of
    temperature. }\label{fig4}
    \end{figure}

According to the simulation, with no externally applied bias the
longitudinal and transverse trapping frequencies for the array of
Fig. 5a are 691~Hz and 35~kHz, respectively. The trap depth is
about 2.7 G. The structure shown here will produce traps only with
in-plane magnetization. For out-of-plane magnetized samples the
design must be changed accordingly. Figure 5b shows an array of
identical patterns on one chip, illustrating another advantage of
hard magnetic atom chips: many patterns can be made on one chip.
The sample shown here will be field tested after it has been
covered with a Au layer to make a mirror.

\section{Discussion}

The success of the foil-based MMOT as well as the magnetic trap
shows that the principle of trapping using permanent magnetic
materials is feasible. Further reduction of the dimensions with
this technique is impossible and the future development will
center on the thin film technology. The thin film design presented
here will be tested in the near future. It should be noted that
this design can still be improved on several points. An important
issue is related to the smoothness of the structures.
Fragmentation of trapped atom clouds close to a current-carrying
wire has recently been attributed to edge roughness of the wires
and/or defects inside the wire \cite{EstAusAsp04,WanLukDem04}. It
is to be expected that the requirements on the edge roughness of
our magnetic film patterns will be similar. The edge roughness in
Fig.5 is on the order of 50 nm, which is a few percent of the
width of the strip. This is much better than some of the first
generation of current-carrying wire atom chips, and only slightly
worse than state-of-the-art wire chips. We expect reduction of the
resist layer will further reduce the roughness. An interesting
alternative to lithographic patterning is to write the patterns
using magneto-optic recording techniques \cite{EriRamHin04}
instead of lithography. This would in principle double the field
gradients and make the patterning process reversible. Yet another
possibility that we will investigate is the use of
electrodeposition on lithographically masked ultrathin conducting
substrates \cite{MLeiThoFah04}. Finally, a promising extension of
our work in the future would be to create atom chips using a
combination of magnetic materials and other elements, such as
current-carrying wires or electrodes for creating electrostatic
potentials.

\section{Summary}

We have used the magnetic properties of FePt foils and thick films
to produce atom chips to trap ultra-cold atoms. FePt patterns were
fabricated from 40 $\mu$m foil using spark-erosion and from 250 nm
film using e-beam lithographic patterning. The patterns generated
in the FePt film have an edge roughness of about 50 nm. The foil
chips were successfully used to trap cold Rb atoms, both as a
mirror magneto-optic trap and as a magnetic trap. It is expected
that the atom chip based on the hard magnetic material will have
certain advantages over chips based on current-carrying wires,
such as the absence of perturbing lead wires, current noise,
thermal noise, and power dissipation. For the two foil chips the
calculated longitudinal and transverse trapping frequencies are
(51~Hz, 6.8~kHz) for the first sample and (34~Hz, 11~kHz) for the
second. For the film chip these frequencies are (690~Hz, 35~kHz)
and (300~Hz, 36~kHz). We  demonstrated that hundreds of traps can
be concentrated per mm$^{2}$ of chip area, with still room for
further optimization. This makes the film chip a promising system
for quantum information processing.
    \begin{figure}[t]
    \centering{\begin{minipage}{60mm}\includegraphics[width=60mm]{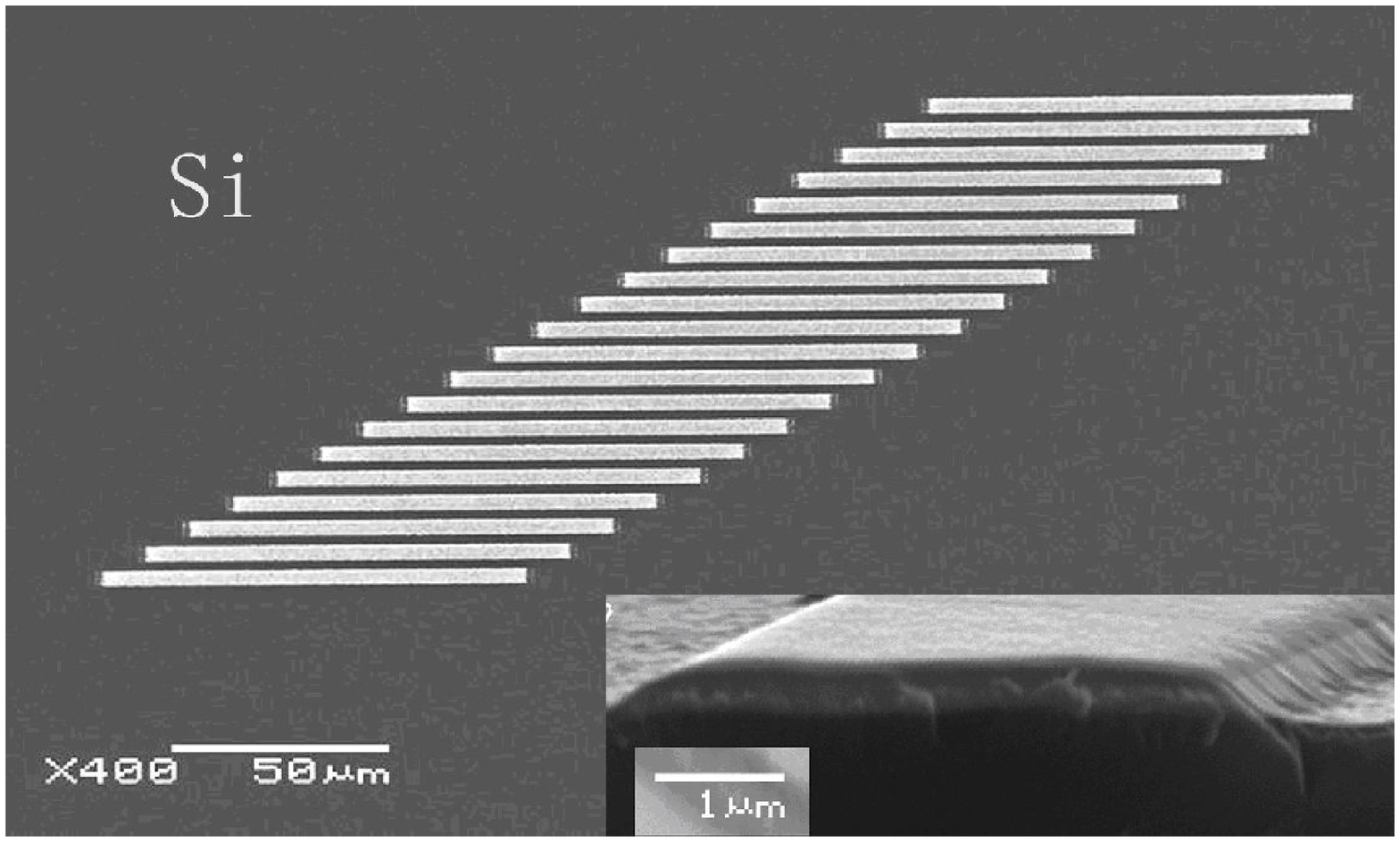}\end{minipage}
    \hspace{20cm}
    \begin{minipage}{60mm}\includegraphics[width=60mm]{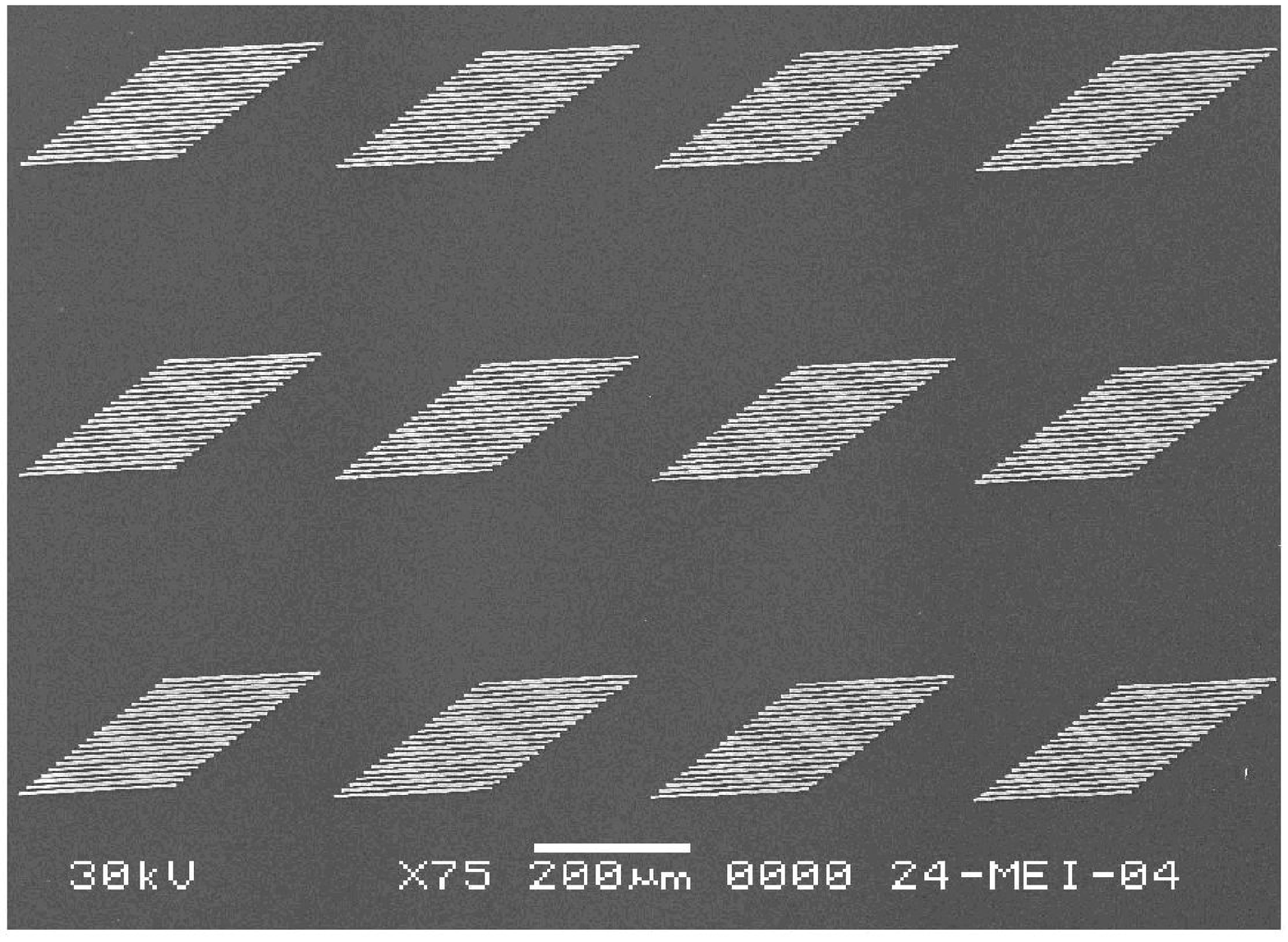}\end{minipage}}
    \caption{(a) Patterned 250 nm FePt film on Si substrate. \\
    The inset shows the cross section of one strip of the \\
    pattern, as well as the edge roughness. The upper layer \\
    consists of SU-8 photoresist. (b) Array of strips on a single
    chip.}\label{fig5}
    \end{figure}

\begin{acknowledgments}

This work was supported by Stichting Fundamenteel Onderzoek der
Materie (FOM), and was made possible by the fabrication and
characterization facilities of the Amsterdam nanoCenter. This work
is part of the research program of the Stichting voor Fundamenteel
Onderzoek van de Materie (Foundation for the Fundamental Research
on Matter) and was made possible by financial support from the
Nederlandse Organisatie voor Wetenschappelijk Onderzoek
(Netherlands Organization for the Advancement of Research). This
work was supported by the EU under contract MRTN-CT-2003-505032.

\end{acknowledgments}

%\bibliography{apssamp}% Produces the bibliography via BibTeX.
\bibliography{amo,mag}

\end{document}